\documentclass[11pt]{article}
\usepackage{graphicx,array,epic,amscd,amssymb,amsmath}
\usepackage[all]{xy}
\usepackage[totalwidth=480pt,totalheight=680pt]{geometry}

\newcommand{\DIS}{\displaystyle}
\newcommand{\qed}{\hfill $\Box$}
\def\C{{\mathbb C}}
\def\Z{{\mathbb Z}}
\def\P{{\mathbb P}}
\def\Q{{\mathbb Q}}
\def\T{{\mathbb T}}
\def\nn{{\nonumber}}
\def\bx{\mbox{\boldmath $x$}}
\def\by{\mbox{\boldmath $y$}}
\def\bI{\mbox{\boldmath $I$}}
\def\bV{\mbox{\boldmath $V$}}
\newtheorem{theorem}{Theorem}
\newtheorem{corollary}{Corollary}
\newtheorem{remark}{Remark}

\newtheorem{proposition}{Proposition}

\begin{document}

\title{\bf Mutations of the cluster algebra of type $\boldsymbol{A^{(1)}_1}$ and\\
the periodic discrete Toda lattice}

\author{Atsushi Nobe\\
\small{Department of Mathematics, Faculty of Education, Chiba University,}\\
\small{1-33 Yayoi-cho Inage-ku, Chiba 263-8522, Japan}\\
\small{nobe@faculty.chiba-u.jp}}
\date{}
\maketitle

\begin{abstract}
A direct connection between two sequences of points, one of which is generated by seed mutations in the cluster algebra of type $A^{(1)}_1$ and the other by time evolutions of the periodic discrete Toda lattice, is explicitly given.
In this construction, each of them is realized as an orbit of a QRT map and specialization of the parameters in the maps and appropriate choices of the initial points relate them.
The connection with the periodic discrete Toda lattice enables us a geometric interpretation of the seed mutations in the cluster algebra of type $A^{(1)}_1$ as addition of points on an elliptic curve arising as the spectral curve of the Toda lattice.
\end{abstract}

\section{Introduction}
A cluster algebra $\mathcal{A}$, which was introduced by Fomin and Zelevinsky in 2002 \cite{FZ02}, is a subalgebra of a field $\mathcal{F}$ isomorphic to the field of rational functions in $n$ independent variables ($n$ is called the rank of $\mathcal{A}$).
Since their introduction cluster algebras have found applications in the field of dynamical systems such as $Y$-systems, discrete soliton equations and discrete Painlev\'e equations \cite{FZ03-2,IIKNS10,IIKKN13,IIKKN13-2,Mase13,Okubo13,Mase16}.
In this paper, we proceed to study links of cluster algebras and discrete integrable systems and establish a direct connection between seed mutations in the cluster algebra of type $A^{(1)}_1$ and time evolutions of the periodic discrete Toda lattice of the lowest dimension through their identifications with QRT maps.
A QRT map is a member of the paradigmatic family of two-dimensional integral maps found by Quispel, Roberts and Thompson in 1989 \cite{QRT89}, which contains many two-dimensional reductions of discrete soliton equations and autonomous limits of the discrete Painelev\'e equations. 
Moreover, a QRT map is geometrically an addition of points on an elliptic curve called the invariant curve of the map \cite{Tsuda04}.
Therefore, the connection with the periodic discrete Toda lattice gives a geometric interpretation of the seed mutations in the cluster algebra via the addition of points on the spectral curve of the Toda lattice.
Since our method to associate a QRT map with a cluster algebra is applicable to any cluster algebra of rank 2, we will list corresponding map dynamical systems to a certain class of cluster algebras of rank 2.

Following \cite{FZ02,FZ03,FZ06}, we briefly review some parts of cluster algebras necessary for our study.
Let $\bx=(x_1,x_2,\ldots,x_n)$ be the set of generators of the ambient field $\mathcal{F}=\mathbb{QP}(\bx)$, where $\P=\left(\P,\cdot,\oplus\right)$ is a semifield endowed with multiplication $\cdot$ and auxiliary addition $\oplus$ and $\mathbb{QP}$ is the group ring of $\P$ over $\Q$.
Also let $\by=(y_1,y_2,\ldots,y_n)$ be an $n$-tuple in $\P^n$ and $B=(b_{ij})$ be an $n\times n$ skew-symmetrizable integral matrix.
The triple $(\bx,\by,B)$ is referred as the (labeled\footnote{In this paper, we consider labeled seeds only. Therefore, we omit the term ``labeled" hereafter.}) seed.
We also refer to $\bx$ as the cluster of the seed, to $\by$ as the coefficient tuple and to $B$ as the exchange matrix.
Elements of $\bx$ and $\by$ are called cluster variables and coefficients, respectively.

If an exchange matrix $B$ is $n\times n$ skew-symmetric then there uniquely corresponds a quiver $Q=(Q_0,Q_1,s,t)$ through the relation
\begin{align*}
b_{ij}=\sharp\left\{\mbox{arrows $i\to j$}\right\}-\sharp\left\{\mbox{arrows $j\to i$}\right\},
\end{align*}
where $Q_0$ is the set of nodes (vertices) labeled by $1,2,\ldots,n$, $Q_1$ is the set of directed edges (arrows) and $s$ (resp. $t$) is the function which maps each arrow to its starting (resp. ending) node. 
We define a path of length $l$ in Q to be a sequence of $l$ arrows $e_1,\ldots,e_l$ with $t(e_i) = s(e_{i+1})$ ($1 \leq i < l$). 
We call a path of length $l$ such that  $t(e_l) = s(e_1)$ an $l$-cycle. 
In particular, $1$-cycle, {\it i.e.}, an arrow $e$ such that $s(e) = t(e)$, is called a loop.
Thus, to each $n\times n$ skew-symmetric matrix with integral entries there uniquely corresponds a quiver $Q$ without loops and $2$-cycles.

We introduce seed mutations.
Let $k\in[1,n]$ be an integer.
The seed mutation $\mu_k$ in the direction $k$ transforms $(\bx,\by,B)$ into the seed $\mu_k(\bx,\by,B)=:(\bx^\prime,\by^\prime,B^\prime)$ defined as follows
\begin{align}
b_{ij}^\prime
&=
\begin{cases}
-b_{ij}&\mbox{$i=k$ or $j=k$},\\
b_{ij}+[-b_{ik}]_+b_{kj}+b_{ik}[b_{kj}]_+&\mbox{otherwise},\\
\end{cases}
\label{eq:mutem}\\
y_j^\prime
&=
\begin{cases}
y_k^{-1}&\mbox{$j=k$},\\
y_jy_k^{[b_{kj}]_+}(y_k\oplus 1)^{-b_{kj}}&\mbox{$j\neq k$},\\
\end{cases}
\label{eq:mutcoef}\\
x_j^\prime
&=
\begin{cases}
\displaystyle\frac{y_k\prod x_i^{[b_{ik}]_+}+\prod x_i^{[-b_{ik}]_+}}{(y_k\oplus 1)x_k}&\mbox{$j=k$},\\
x_j&\mbox{$j\neq k$},\\
\end{cases}
\label{eq:mutcv}
\end{align}
where we define $[a]_+:=\max(a,0)$ for $a\in\Z$.

Let $\T_n$ be the $n$-regular tree whose edges are labeled by $1, 2, \ldots, n$ so that the $n$ edges emanating from each vertex receive different labels.
We write $t\ \overset{k}{\begin{xy}\ar @{-}(10,0)\end{xy}}\ t^\prime$ to indicate that vertices $t,t^\prime\in\T_n$ are joined by an edge labeled by $k$.
We assign a seed $\Sigma_t=(\bx_t,\by_t,B_t)$ to every vertex $t\in\T_n$ so that the seeds assigned to the endpoints of any edge $t\ \overset{k}{\begin{xy}\ar @{-}(10,0)\end{xy}}\ t^\prime$ are obtained from each other by the seed mutation in direction $k$.
We refer the assignment $\T_n\ni t\mapsto\Sigma_t$ to a cluster pattern.
We write the elements of $\Sigma_t$ as follows
\begin{align*}
\bx_t=(x_{1;t},\ldots,x_{n;t}),\qquad
\by_t=(y_{1;t},\ldots,y_{n;t}),\qquad
B_t=(b_{ij}^t).
\end{align*}

Given a cluster pattern $\T_n\ni t\mapsto\Sigma_t$, we denote the union of clusters of all seeds in the pattern by
\begin{align*}
\mathcal{X}
=
\bigcup_{t\in\T_n}\bx_t
=
\left\{
x_{i;t}\ |\ t\in\T_n,\ 1\leq i\leq n
\right\}.
\end{align*}
The cluster algebra $\mathcal{A}$ associated with a given cluster pattern is the $\mathbb{ZP}$-subalgebra of the ambient field $\mathcal{F}$ generated by all cluster variables: $\mathcal{A}=\mathbb{ZP}[\mathcal{X}]$.

Let ${\rm Trop}(u_1,u_2,\ldots,u_m)$ be an abelian multiplicative group freely generated by $u_1,u_2,\ldots,u_m$.
We define the auxiliary addition $\oplus$ in ${\rm Trop}(u_1,u_2,\ldots,u_m)$ by
\begin{align*}
\prod_ju_j^{a_j}\oplus\prod_ju_j^{b_j}=\prod_ju_j^{\min(a_j,b_j)}.
\end{align*}
We call $\left({\rm Trop}(u_1,u_2,\ldots,u_m),\cdot,\oplus\right)$ a tropical semifield.
We say that a cluster pattern $t\mapsto\Sigma_t$ on $\T_n$ or the corresponding cluster algebra $\mathcal{A}$ has principal (tropical) coefficients at a vertex $t_0$ if $\P={\rm Trop}(y_1,y_2,\ldots,y_n)$ and $\by_{t_0}=(y_1,y_2,\ldots,y_n)$.

This paper is organized as follows.
In section \ref{sec:TLQRT}, we introduce the periodic discrete Toda lattice of dimension 4 and rewrite it as a QRT map $\varphi_{\rm TL}:\P^1(\C)\times\P^1(\C)\to\P^1(\C)\times\P^1(\C)$.
In this formulation, we use the additive group structure of an elliptic curve which is equivalent to the time evolution of the Toda lattice.
In section \ref{sec:CAQRT}, we show that the composition $\mu_2\circ\mu_1$ of the seed mutations $\mu_1$ and $\mu_2$ of the cluster algebra of type $A^{(1)}_1$ induces a QRT map  $\varphi_{\rm CA}:\P^1(\C)\times\P^1(\C)\to\P^1(\C)\times\P^1(\C)$.
In section \ref{sec:TLCA}, we give the main theorem concerning a direct connection between two QRT maps $\varphi_{\rm TL}$ and $\varphi_{\rm CA}$.
By using the theorem, we relate special solutions to the periodic discrete Toda lattice with the cluster variables of type $A^{(1)}_1$ obtained from a certain initial seed.
Section \ref{sec:CONCL} is devoted to concluding remarks.
Finally, in appendix \ref{asec:CAR2QRT}, we extend the correspondence between the cluster algebra of type $A^{(1)}_1$ and the QRT map and list corresponding map dynamical systems to all cluster algebras of rank 2 associated with finite or affine Lie algebras.

\section{A QRT map reduced from the periodic discrete Toda lattice}
\label{sec:TLQRT}
Let us consider a four-dimensional map $\chi_1:\C^{4}\to\C^{4}; (\bI,\bV)\mapsto (\bar{\bI,}\bar{\bV})$, where we put
\begin{align*}
\bI=\left(I_1,I_{2}\right),\qquad
\bV=\left(V_1,V_{2}\right)
\end{align*}
and the evolution is defined by
\begin{align}
\begin{cases}
\DIS\bar I_i+\bar V_{i-1}=I_i+V_i,\\
\DIS\bar I_i \bar V _i=I_{i+1}V_i,\\
\end{cases}
\label{eq:TETL}
\end{align}
for $i=1,2$. 
Here the subscripts are reduced modulo 2.
This map is known as the periodic discrete Toda lattice of the lowest dimension \cite{HTI93}.
One can solve the equation (\ref{eq:TETL}) for $\bar I_i$ and $\bar V_i$ as follows
\begin{align}
\begin{cases}
\DIS\bar I_i=\frac{\DIS I_i+V_i}{\DIS I_{i-1}+V_{i-1}}I_{i-1},\\[10pt]
\DIS\bar V_i=\frac{\DIS I_{i+1}+V_{i+1}}{\DIS I_{i}+V_{i}}V_{i}.\\
\end{cases}
\label{eq:TETL2}
\end{align}
Then we find that this birational map is nothing but the discrete Toda lattice of type $A^{(1)}_{1}$ \cite{Suris03} firstly introduced by Hirota in 1977 \cite{Hirota77}.
Therefore, we also refer to the map $\chi_1$ as the discrete Toda lattice of type $A^{(1)}_1$.

Let $2\times2$ matrices $L$ and $M$ be
\begin{align*}
L=\left(
\begin{matrix}
I_2+V_1&1-I_1V_1/y\\
I_2V_2-y&I_1+V_2\\
\end{matrix}\right),\qquad
M=\left(
\begin{matrix}
I_2&1\\
-y&I_1\\
\end{matrix}\right),
\end{align*}
where $y$ is the spectral parameter.
Then the Lax form of the Toda lattice $\chi_1$ is given by
\begin{align*}
\bar LM=ML.
\end{align*}
By using the Lax matrix $L$, the spectral curve $\gamma_1$ of the Toda lattice $\chi_1$ is defined to be the following compact curve on the projective plane $\P^2(\C)$
\begin{align*}
\gamma_1
&:=
\left(f(x,y)=0\right)\cup\left\{P_\infty,P_\infty^\prime\right\}\subset\P^2(\C),\\
f(x,y)&:=y\det\left(L+xE\right),
\end{align*}
where $E$ is the $2\times2$ identity matrix and $P_\infty, P_\infty^\prime$ are the points at infinity.
Upon introduction of the homogeneous coordinate of $\P^2(\C)$, $(x,y)\mapsto[X:Y:Z]=[x:y:1]$, we have $P_\infty=[1:0:0]$ and $P_\infty^\prime=[0:1:0]$.
For generic choice of $\bI$ and $\bV$, the curve $\gamma_1$ is an elliptic curve.
The defining polynomial $f(x,y)$ of $\gamma_1$ is expanded as follows
\begin{align*}
&f(x,y)=y^2+y\left(x^{2}+c_1x+c_0\right)+c_{-1},\\
&c_1=I_1+I_2+V_1+V_2,\qquad
c_0=I_1I_2+V_1V_2,\qquad
c_{-1}=I_1I_2V_1V_2.
\end{align*}
The coefficients $c_{1},c_0,c_{-1}$ of $f(x,y)$ are the conserved quantities of the Toda lattice $\chi_1$.

Put
\begin{align*}
\varphi_1=I_2V_2-y,\qquad
\varphi_2=x+I_1+V_2.
\end{align*}
Then $\varphi(x,y)={}^t\left(\varphi_1,\varphi_2\right)$ is the eigenvector of the matrix $L$ associated with the eigenvalue $-x$, {\it i.e.}, the following holds
\begin{align*}
\left(L+xE\right)\varphi(x,y)=0.
\end{align*}
We see that the solution
\begin{align*}
(x,y)=\left(-I_1-V_2,I_2V_2\right)
\end{align*}
to the system of equations $\varphi_1=0$ and $\varphi_2=0$ uniquely determines the point $P=(x,y)$ on the spectral curve $\gamma_1$.
The map $\C^4\to\P^2(\C);(\bI,\bV)\mapsto(x,y)=\left(-I_1-V_2,I_2V_2\right)$ is called the eigenvector map.

Let the initial value of the map $\chi_1$ be $\left(\bI^0,\bV^0\right)\in\C^4$.
We refer to $\left(\bI^t,\bV^t\right)$ ($t\geq0$) as the $t$-th iteration of the map $\chi_1$ from $\left(\bI^0,\bV^0\right)$:
\begin{align*}
\left(\bI^t,\bV^t\right)=\underbrace{\chi_1\circ\chi_1\circ\cdots\circ\chi_1}_{t}\left(\bI^0,\bV^0\right).
\end{align*}
For $t\geq0$, let $P^t$ be the point on the curve $\gamma_1$ given by the eigenvector map:
\begin{align*}
P^t=(x^t,y^t)=\left(-I_1^t-V_2^t,I_2^tV_2^t\right).
\end{align*}
Then the sequence $\left\{P^t\right\}_{t\geq0}$ of points on the elliptic curve $\gamma_1$ is generated by successive applications of the map $\chi_1$ to the initial point $P^0$.

On the other hand, it is known that the time evolution $P^t\mapsto P^{t+1}$ of the discrete Toda lattice $\chi_1$ of type $A^{(1)}_1$ is an addition of points on the elliptic curve $\gamma_1$ arising as its spectral curve \cite{Iwao08,Nobe13}.
\begin{theorem}\label{thm:paTL}
Let $T$ be the point $\left(-c_1,b\right)$ on the elliptic curve $\gamma_1$, where $b=-V_1^0V_2^0$.
Then we have
\begin{align*}
P^{t+1}=P^{t}+T
\end{align*}
for $t\geq0$.
Here we choose the point $P_\infty=[1:0:0]$ at infinity as the unit of addition.
\qed
\end{theorem}

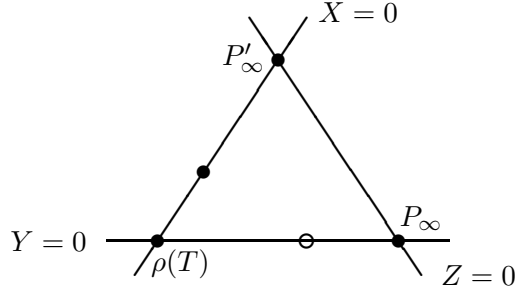
\begin{figure}[htbp]
\centering
\unitlength=.03in
\def\arraystretch{1.0}
\begin{picture}(80,60)(-15,-10)
\thicklines
\thicklines
\put(-5,0){\line(1,0){60}}
\put(0,-6){\line(2,3){30}}
\put(50,-6){\line(-2,3){30}}
\put(4,0){\circle*{2.5}}
\put(12,12){\circle*{2.5}}
\put(30,0){\circle{2.5}}
\put(46,0){\circle*{2.5}}
\put(25,31.5){\circle*{2.5}}
\put(39,40){\makebox(0,0){$X=0$}}
\put(-15,0){\makebox(0,0){$Y=0$}}
\put(60,-6){\makebox(0,0){$Z=0$}}
\put(8,-4){\makebox(0,0){$\rho(T)$}}
\put(50,4){\makebox(0,0){$P_\infty$}}
\put(19,32){\makebox(0,0){$P_\infty^\prime$}}
\end{picture}
\caption{
Configuration of the intersection points of the curve $\gamma_1$ and the coordinate lines on the plane $\P^2(\C)$.
The base points of the pencil $\left\{\gamma_1\right\}_{c_1\in\P^1(\C)}$ are denoted by the filled circles.
}
\label{fig:P2gamma}
\end{figure}

Now we transform the discrete Toda lattice $\chi_1$ of type $A^{(1)}_1$ into a QRT map in terms of the additive group structure of the elliptic curve $\gamma_1$ \cite{Tsuda04}.
Let us shift the point $T=\left(-c_1,b\right)$ to the origin $(x,y)=(0,0)$ of the plane $\P^2(\C)$.
By employing the transformation $\rho:(x,y)\mapsto(x+c_1,y-b)$ on $\C^2$, we have $\rho(T)=(0,0)$ and 
\begin{align*}
\left(f\circ\rho^{-1}\right)(x,y)={y}^2+y\left({x}^2-c_1x+a\right)+bx\left(x-c_1\right),
\end{align*}
where we put 
\begin{align}
&a
=
\sqrt{c_0^2-4c_{-1}}
=
I_1^0I_2^0-V_1^0V_2^0,
\qquad
b
=
\frac{a-c_0}{2}
=
\frac{\sqrt{c_0^2-4c_{-1}}-c_0}{2}
=
-V_1^0V_2^0,
\label{eq:params}
\end{align}
which are the conserved quantity of the map $\chi_1$.
The point $\rho(T)$ and the curve $\left(f\circ\rho^{-1}\right)(x,y)$ are respectively given by the homogeneous coordinate $(x,y)\mapsto[X:Y:Z]=[x:y:1]$ as
\begin{align*}
\rho(T)&\mapsto[0:0:1],\\
\left(f\circ\rho^{-1}\right)(x,y)&\mapsto Y^2Z+Y\left(X^2-c_1XZ+aZ^2\right)+bXZ\left(X-c_1Z\right).
\end{align*}
Thus, the curve
\begin{align*}
\widetilde\gamma_1:=\left(\left(f\circ\rho^{-1}\right)(x,y)=0\right)\cup\left\{P_\infty,P_\infty^\prime\right\}
\end{align*}
passes through the three intersection points $\rho(T)$, $P_\infty$ and $P_\infty^\prime$ of the axes $X=0$, $Y=0$ and $Z=0$ of the plane $\P^2(\C)$ (see figure \ref{fig:P2gamma}).
The intersection multiplicities of $\widetilde\gamma_1$ and $\left\{X=0\right\}\cup\left\{Y=0\right\}\cup\left\{Z=0\right\}$ at $\rho(T)$, $P_\infty$ and $P_\infty^\prime$ are two, two and three, respectively.

By employing the change of coordinate
\begin{align*}
\sigma:\P^2(\C)\to\P^2(\C); [X:Y:Z]\mapsto[U:V:W]=[Z:X:Y],
\end{align*}
we obtain
\begin{align*}
\left(\sigma\circ\rho\right)(T)=[1:0:0]=:\mathcal{T},\qquad
\sigma\left(P_\infty\right)=[0:1:0]=:\mathcal{O},\qquad
\sigma\left(P_\infty^\prime\right)=[0:0:1].
\end{align*}
In this coordinate, $\widetilde\gamma_1$ has the form
\begin{align}
\widetilde\gamma_1=\left(bxy^2+y^2-bc_1x^2y-c_1xy+ax^2+x=0\right)\cup\left\{\mathcal{O},\mathcal{T},\mathcal{S}\right\},
\label{eq:spectralcurve}
\end{align}
where $\mathcal{S}:=[1:c_1:0]$ and we use the inhomogeneous coordinate $[U:V:W]\mapsto(x,y)=\left({U}/{W},{V}/{W}\right)$.
We find that the sequence $\left\{P^t\right\}_{t\geq0}$ of points on $\gamma_1$ is mapped into the sequence $\{\widetilde{P}^t\}_{t\geq0}$ of points on $\widetilde\gamma_1$, where we put
\begin{align}
\widetilde{P}^t=\left(\frac{1}{V_2^t\left(I_2^t+V_1^t\right)},\frac{1}{V_2^t}\right).
\label{eq:pointP}
\end{align}
We then obtain the following corollary to theorem \ref{thm:paTL}.
\begin{corollary}\label{col:paTLQRT}
Suppose that $\mathcal{O}=[0:1:0]$ is the unit of addition of points on the elliptic curve $\widetilde\gamma_1$.
Let $\mathcal{T}$ be the point $[1:0:0]$ on $\widetilde\gamma_1$.
Then we have
\begin{align}
\widetilde P^{t+1}=\widetilde P^{t}+\mathcal{T}
\label{eq:addontgam}
\end{align}
for $t\geq0$.
\qed
\end{corollary}

The time evolution $\widetilde P^t\mapsto\widetilde P^{t+1}$ of points on $\widetilde\gamma$ reduced from the discrete Toda lattice $\chi_1$ of type $A^{(1)}_1$ is geometrically realized in terms of the intersection of the elliptic curve $\widetilde\gamma_1$ and two lines $\ell_1$ and $\ell_2$ in the following manner.
Let the line passing through $\widetilde P^t$ and parallel to the axis $y=0$ be $\ell_1$:
\begin{align*}
\ell_1=\left(V_2^ty-1=0\right).
\end{align*}
Then $\ell_1$ intersects $\widetilde\gamma_1$ at another point $\widetilde Q^t$:
\begin{align*}
\widetilde{Q}^t=\left(\frac{1}{V_2^t\left(I_1^t+V_1^t\right)},\frac{1}{V_2^t}\right).
\end{align*}
Also let the line passing through $\widetilde Q^t$ and parallel to the axis $x=0$ be $\ell_2$:
\begin{align*}
\ell_2=\left(V_2^t\left(I_1^t+V_1^t\right)x-1=0\right).
\end{align*}
Then $\ell_2$ intersects $\widetilde\gamma_1$ at
\begin{align*}
\left(\frac{1}{V_2^t\left(I_1^t+V_1^t\right)},\frac{I_2^t+V_2^t}{V_2^t\left(I_1^t+V_1^t\right)}\right)
=
\left(\frac{1}{V_2^{t+1}\left(I_2^{t+1}+V_1^{t+1}\right)},\frac{1}{V_2^{t+1}}\right)
=
\widetilde P^{t+1}.
\end{align*}
Thus, the intersection points of the curve $\widetilde\gamma_1$ and the line $\ell_1$ are $\widetilde P^t$ and $\widetilde Q^t$; and the intersection points of $\widetilde\gamma_1$ and $\ell_2$ are $\widetilde Q^t$ and $\widetilde P^{t+1}$.
This is a geometric realization of the time evolution $\widetilde P^t\mapsto\widetilde P^{t+1}$ of the discrete Toda lattice $\chi_1$ of type $A^{(1)}_1$. 

Since the time evolution $\widetilde P^t\mapsto\widetilde P^{t+1}$ is originated in the additive group structure of the elliptic curve $\widetilde\gamma_1$ as shown above, it gives a QRT map on the curve $\widetilde\gamma_1$.
Actually, we observe that the map $\widetilde P^t\mapsto\widetilde P^{t+1}$ coincides with the QRT map $\varphi_{\rm TL}$ given by the following matrices \cite{QRT89}
\begin{align}
A_{\rm TL}
=
\left(\begin{matrix}
0&0&a\\
b&0&1\\
1&0&0\\
\end{matrix}\right),\qquad
B_{\rm TL}
=
\left(\begin{matrix}
0&b&0\\
0&1&0\\
0&0&0\\
\end{matrix}\right).
\label{eq:TodaQRTmat}
\end{align}
Let $f_i$ and $g_i$ ($i=1,2,3$) be
\begin{align*}
\left(
\begin{matrix}
f_1\\ f_2\\ f_3\\
\end{matrix}
\right)
&=
A_{\rm TL}\left(
\begin{matrix}
\left({y^t}\right)^2\\ y^t\\ 1\\
\end{matrix}
\right)
\times
B_{\rm TL}\left(
\begin{matrix}
\left({y^t}\right)^2\\ y^t\\ 1\\
\end{matrix}
\right),\\
\left(
\begin{matrix}
g_1\\ g_2\\ g_3\\
\end{matrix}
\right)
&=
{}^tA_{\rm TL}\left(
\begin{matrix}
\left({x^{t+1}}\right)^2\\ x^{t+1}\\ 1\\
\end{matrix}
\right)
\times
{}^tB_{\rm TL}\left(
\begin{matrix}
\left({x^{t+1}}\right)^2\\ x^{t+1}\\ 1\\
\end{matrix}
\right).
\end{align*}
We then obtain the QRT map $\varphi_{\rm TL}:\P^1(\C)\times\P^1(\C)\to\P^1(\C)\times\P^1(\C);(x^t,y^t)\mapsto(x^{t+1},y^{t+1})$ as follows
\begin{align*}
x^{t+1}
&=
\frac{f_1-f_2 x^t}{f_2-f_3 x^t}
=
\frac{-\left(bx^{t}+1\right)\left(y^{t}\right)^2}{b\left(bx^{t}+1\right)\left(y^{t}\right)^2-(a-b)x^{t}},\\
y^{t+1}
&=
\frac{g_1-g_2 y^t}{g_2-g_3 y^t}
=
\frac{a\left(x^{t+1}\right)^2+x^{t+1}}{\left(bx^{t+1}+1\right)y^{t}}.
\end{align*}

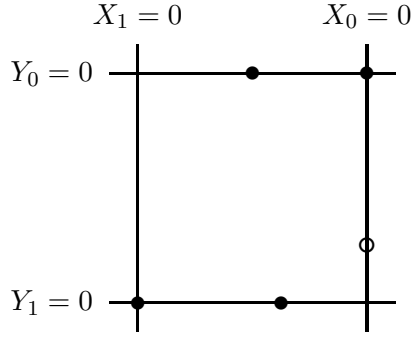
\begin{figure}[htbp]
\centering
\unitlength=.03in
\def\arraystretch{1.0}
\begin{picture}(80,60)(-15,-5)
\thicklines
\put(-5,0){\line(1,0){50}}
\put(-5,40){\line(1,0){50}}
\put(0,-5){\line(0,1){50}}
\put(40,-5){\line(0,1){50}}
\put(0,0){\circle*{2.5}}
\put(20,40){\circle*{2.5}}
\put(40,40){\circle*{2.5}}
\put(40,10){\circle{2.5}}
\put(25,0){\circle*{2.5}}
\put(-15,40){\makebox(0,0){$Y_0=0$}}
\put(-15,0){\makebox(0,0){$Y_1=0$}}
\put(0,50){\makebox(0,0){$X_1=0$}}
\put(40,50){\makebox(0,0){$X_0=0$}}
\end{picture}
\caption{
Configuration of the intersection points of the curve $\widetilde\gamma$ and the coordinate lines on the plane $\P^1(\C)\times\P^1(\C)$.
The base points of the pencil $\left\{\widetilde\gamma_1\right\}_{\lambda\in\P^1(\C)}$ are denoted by the filled circles.
}
\label{fig:P1P1gamma}
\end{figure}

We also obtain the invariant curve of the QRT map $\varphi_{\rm TL}$ as
\begin{align}
(x^2\ x\ 1)
\left(A_{\rm TL}+\lambda B_{\rm TL}\right)
\left(\begin{matrix}
y^2\\ y\\1
\end{matrix}
\right)
=bxy^2+y^2+b\lambda x^2y+\lambda xy+ax^2+x=0,
\label{eq:invariantcurve}
\end{align}
where $\lambda$ is the conserved quantity of $\varphi_{\rm TL}$.
Note that this is the affine part of the elliptic curve $\widetilde\gamma_1$ given by (\ref{eq:spectralcurve}) with imposing $c_1=-\lambda$.
The compactification of the invariant curve (\ref{eq:invariantcurve}) on $\P^1(\C)\times\P^1(\C)$ is
\begin{align*}
\left(bxy^2+y^2+b\lambda x^2y+\lambda xy+ax^2+x=0\right)
\cup
\left\{
\left([0:1],[0:1]\right),
\left([0:1],[-b\lambda:a]\right),
\left([-b:1],[0:1]\right)
\right\}
\end{align*}
in the homogeneous coordinate $(x,y)\mapsto\left([X_0:X_1],[Y_0:Y_1]\right)=([1:x],[1:y])$ (see figure \ref{fig:P1P1gamma}).
We also refer to $\widetilde\gamma_1$ as this compact curve.

Let $\widetilde P^t=(x^t,y^t)$ be a point on the curve $\widetilde\gamma_1$.
We then have
\begin{align}
x^t
=
\frac{1}{V_2^t\left(I_2^t+V_1^t\right)},\qquad
y^t
=
\frac{1}{V_2^t}
\label{eq:xyIV}
\end{align}
from (\ref{eq:pointP}).
The equivalence of the maps $\chi_1$ and $\varphi_{\rm TL}$ is shown as follows.

Given initial values $\bI^0,\bV^0$ of $\chi_1$, the values of the parameters $a$ and $b$ of $\varphi_{\rm TL}$ are fixed (see (\ref{eq:params})):
\begin{align*}
a=I_1^0I_2^0-V_1^0V_2^0,\qquad
b=-V_1^0V_2^0.
\end{align*}
The initial point $(x^0,y^0)$ on the plane $\P^1(\C)\times\P^1(\C)$ is determined by (\ref{eq:xyIV}) with imposing $t=0$.
Note that the initial values $\bI^0,\bV^0$ of $\chi_1$ must be chosen so that the initial point $(x^0,y^0)$ is not a base point of the pencil $\left\{\widetilde\gamma\right\}_{\lambda\in\P^1(\C)}$.
The base points are the following four points in the homogeneous coordinate (see figure \ref{fig:P1P1gamma})
\begin{align}
\left([1:0],[1:0]\right),
\qquad
\left([-a:1],[1:0]\right),
\qquad
\left([0:1],[0:1]\right),
\qquad
\left([-b:1],[0:1]\right).
\label{eq:bpp1p1}
\end{align}
Then $\varphi_{\rm TL}$ generates a sequence of points starting from $(x^0,y^0)$ on the invariant curve (\ref{eq:invariantcurve}).
Conversely, if the values of the parameters $a, b$ and the initial point $(x^0,y^0)$ of $\varphi_{\rm TL}$ are given then we obtain the initial values $\bI^0,\bV^0$ of $\chi_1$ through
\begin{align}
I_1^t
=
\frac{\left(a-b\right)x^t}{\left(bx^t+1\right)y^t},\qquad
I_2^t
=
\frac{\left(bx^t+1\right)y^t}{x^t},\qquad
V_1^t
=-by^t,\qquad
V_2^t
=
\frac{1}{y^t}
\label{eq:fromxytoIV}
\end{align}
with imposing $t=0$.
Then $\chi_1$ generates a sequences of points starting from $\bI^0,\bV^0$ on $\C^4$ since $(x^0,y^0)$ is not a base point of the pencil $\left\{\widetilde\gamma\right\}_{\lambda\in\P^1(\C)}$ (see (\ref{eq:bpp1p1})).
Thus, we identify the map $\chi_1$ with the map $\varphi_{\rm TL}$.

Note that the conserved quantity $\lambda$ of the QRT map $\varphi_{\rm TL}$ is recovered from the dependent variables $\bI^t$ and $\bV^t$ of the Toda lattice $\chi_1$ via its conserved quantity $c_1$:
\begin{align}
\lambda
&=
-c_1
\nn\\
&=
-I_1^t-I_2^t-V_1^t-V_2^t
\nn\\
&=
-\frac{\left(a-b\right)x^t}{\left(bx^t+1\right)y^t}-\frac{\left(bx^t+1\right)y^t}{x^t}+by^t-\frac{1}{y^t}.
\label{eq:conservedq}
\end{align}
We complete this in the following proposition.
\begin{proposition}
The time evolution of the discrete Toda lattice $\chi_1$ of type $A^{(1)}_1$ on the spectral curve $\widetilde\gamma_1$ is realized as the QRT map $\varphi_{\rm TL}$ on it given by the matrices (\ref{eq:TodaQRTmat}) with imposing (\ref{eq:params}).
The conserved quantity of $\varphi_{\rm TL}$ is given by (\ref{eq:conservedq}).
\qed
\end{proposition}

\section{Seed mutations of type $\boldsymbol{A^{(1)}_1}$ and a QRT map}
\label{sec:CAQRT}
\begin{figure}[htbp]
\center
$
\xymatrix{{t_{-2}}\ar @{-}(3,-2);(13,-12)^1&&t_0&\ar @{-}(33,-2);(43,-12)^1&t_2&\ar @{-}(60,-2);(70,-12)^1&\\
\ar @{--}(-14,-12);(-4,-2)&{t_{-1}}\ar @{-}(18,-12);(28,-2)^2&\ar @{-}(46,-12);(56,-2)^2&t_1&\ar @{--}(73,-12);(83,-2)&t_3\\}
$
\caption{
The regular binary tree $\T_2$ and the cluster pattern $\T_2\ni t_m\mapsto\Sigma_m$.
}
\label{fig:binarytree}
\end{figure}
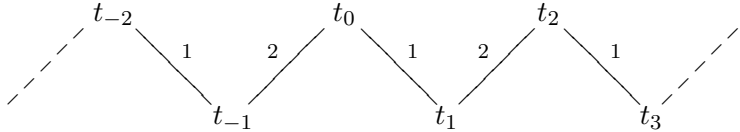

Now we introduce the cluster algebra $\mathcal{A}$ of type $A^{(1)}_1$ which is of rank 2 and has principal coefficients.
Let the initial seed $\Sigma_0=\left(\bx_0,\by_0,B_0\right)$ be
\begin{align}
\bx_0=\left(x_1,x_2\right),\qquad
\by_0=\left(y_1,y_2\right),\qquad
B_0=\left(\begin{matrix}0&-2\\2&0\\\end{matrix}\right),
\label{eq:seedA11}
\end{align}
where $\bx_0$ is the cluster, $\by_0$ is the coefficient tuple and $B_0$ is the exchange matrix. 
Also let the semifield of coefficients be $\P={\rm Trop}\mkern2mu(y_1,y_2)$, the tropical semifield generated by $y_1$ and $y_2$.
We consider the regular binary tree $\T_2$ whose edges are labeled by the numbers 1 and 2. 
The tree $\T_2$ is an infinite chain (see figure \ref{fig:binarytree}).

\begin{figure}[htbp]
\centering
\unitlength=.06in
\def\arraystretch{1.0}
\begin{picture}(30,5)(0,0)
\thicklines
\put(1,.5){\circle{2}}
\put(10,.5){\circle{2}}
\put(1.8,1){\vector(1,0){7.2}}
\put(1.8,0){\vector(1,0){7.2}}
\put(2.4,1){\vector(-1,0){.5}}
\put(2.4,0){\vector(-1,0){.5}}
\put(1,3){\makebox(0,0){$1$}}
\put(10,3){\makebox(0,0){$2$}}
\put(21,.5){\circle{2}}
\put(30,.5){\circle{2}}
\put(29.1,1){\vector(-1,0){7.2}}
\put(29.1,0){\vector(-1,0){7.2}}
\put(21,3){\makebox(0,0){$1$}}
\put(30,3){\makebox(0,0){$2$}}
\end{picture}
\caption{
The Dynkin diagram of type $A^{(1)}_1$ (left) and the quiver associated with the exchange matrix $B_0$ (right).
}
\label{fig:Dynkin}
\end{figure}
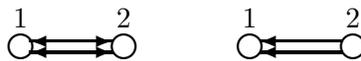

Let $\T_2\ni t_m\mapsto\Sigma_m=\left(\bx_m,\by_m,B_m\right)$ be the cluster pattern of the cluster algebra $\mathcal{A}$ given in figure \ref{fig:binarytree}.
Hereafter, we fix the cluster pattern.
We see from (\ref{eq:mutem}) that the following holds
\begin{align*}
B_m
=
\begin{cases}
B_0&\mbox{$m$ even,}\\
-B_0&\mbox{$m$ odd.}\\
\end{cases}
\end{align*}
Since the Cartan counterpart $A(B_m)$ \cite{FZ03} of the exchange matrix $B_m$ for any $m\in\Z$ is the generalized Cartan matrix of type $A^{(1)}_1$:
\begin{align*}
A(B_m)
=
\left(2\delta_{ij}-\left|b_{ij}\right|\right)
=
\left(\begin{matrix}2&-2\\-2&2\\\end{matrix}\right),
\end{align*}
we refer to the cluster algebra $\mathcal{A}$ as of type $A^{(1)}_1$.
The Dynkin diagram of type $A^{(1)}_1$ and the quiver associated with the exchange matrix $B_0$ are given in figure \ref{fig:Dynkin}.
The seed mutations $\mu_1$ and $\mu_2$ are also referred as of type $A^{(1)}_1$.

By applying (\ref{eq:mutcoef}) repeatedly, we find that the coefficients $y_{1;m}$ and $y_{2;m}$ for $m\geq2$ are the following monomials in the initial coefficients $y_1$ and $y_2$, respectively:
\begin{align*}
y_{1;m}
&=
\begin{cases}
\DIS\left(y_{1}\right)^{-(m-1)}\left(y_{2}\right)^{-(m-2)}&\mbox{$m$ even,}\\
\DIS\left(y_{1}\right)^{m-2}\left(y_{2}\right)^{m-3}&\mbox{$m$ odd,}\\
\end{cases}\\
y_{2;m}
&=
\begin{cases}
\DIS\left(y_{1}\right)^{m-2}\left(y_{2}\right)^{m-3}&\mbox{$m$ even,}\\
\DIS\left(y_{1}\right)^{-(m-1)}\left(y_{2}\right)^{-(m-2)}&\mbox{$m$ odd.}\\
\end{cases}
\end{align*}

We now consider the map 
\begin{align*}
\bx_m=(x_{1;m},x_{2;m})
\overset{\mu_1}{\longmapsto}
\bx_{m+1}=(x_{1;m+1},x_{2;m+1})
\overset{\mu_2}{\longmapsto}
\bx_{m+2}=(x_{1;m+2},x_{2;m+2})
\end{align*}
generated by successive application of the seed mutations $\mu_1$ and $\mu_2$.
We then find that the first map $\mu_1$ fixes $x_{2;m}$; and the second map $\mu_2$ fixes $x_{1;m+1}$.
This is the same property as the QRT map (see \cite{Tsuda04}).
Therefore, we can realize the composition $\mu_2\circ\mu_1$ of the seed mutations $\mu_1$ and $\mu_2$ of type $A^{(1)}_1$ as a QRT map on $\P^1(\C)\times\P^1(\C)$.

\begin{proposition}
Let $\T_2\ni t_m\mapsto\Sigma_m$ be the cluster pattern of the cluster algebra $\mathcal{A}$ of type $A^{(1)}_1$.
The composition $\mu_2\circ\mu_1$ of the seed mutations $\mu_1$ and $\mu_2$ in $\mathcal{A}$ induces the QRT map $\varphi_{\rm CA}$ given by the matrices
\begin{align}
A_{\rm CA}
=
\left(\begin{matrix}
0&0&y_1^2y_2^3\\
0&0&0\\
y_1y_2^2&0&1\\
\end{matrix}\right),\qquad
B_{\rm CA}
=
\left(\begin{matrix}
0&0&0\\
0&1&0\\
0&0&0\\
\end{matrix}\right).
\label{eq:CAQRTmat}
\end{align}
Moreover, the invariant curve of $\varphi_{\rm CA}$ is given as
\begin{align}
y_1y_2^2w^2+\lambda zw+y_1^2y_2^3z^2+1=0.
\label{eq:CAQRTinvari}
\end{align}
\end{proposition}

(Proof)\qquad
Let us introduce new variables $z^t$ and $w^t$ which are associated with the seed $\Sigma_{2t}=(\bx_{2t},\by_{2t},B_{2t})$ of the cluster algebra $\mathcal{A}$ as
\begin{align*}
&z^t:=
\begin{cases}
\DIS\frac{x_{1;2t}}{(y_1y_2)^t}&(t\geq1),
\\[10pt]
\DIS\frac{x_1}{y_1(y_2)^2}&(t=0),
\end{cases}
\qquad
w^t:=
\begin{cases}
\DIS\frac{x_{2;2t}}{(y_1y_2)^t}&(t\geq1), 
\\[10pt]
\DIS\frac{x_2}{y_2}&(t=0).
\end{cases}
\end{align*}
Then the map $\mu_2\circ\mu_1:(x_{1;m},x_{2;m})\mapsto(x_{1;m+2},x_{2;m+2})$, which is explicitly given by
\begin{align*}
&x_{1;m+2}
=
\begin{cases}
\DIS\frac{\left(x_{2;m}\right)^2+y_1^{m-1}y_2^{m-2}}{x_{1;m}}&(m\geq2),
\\[10pt]
\DIS\frac{y_1(x_2)^2+1}{x_1}&(m=0),
\end{cases}
\\
&x_{2;m+2}
=
\begin{cases}
\DIS\frac{\left(x_{1;m+2}\right)^2+y_1^{m}y_2^{m-1}}{x_{2;m}}&(m\geq2),
\\[10pt]
\DIS\frac{y_2\left(x_{1;2}\right)^2+1}{x_2}&(m=0),
\end{cases}
\end{align*}
reduces to $(z^t,w^t)\mapsto(z^{t+1},w^{t+1})$;
\begin{align*}
z^{t+1}
=
\frac{y_1y_2^2\left(w^t\right)^2+1}{y_1^2y_2^3z^t},\qquad
w^{t+1}
=
\frac{y_1^2y_2^3\left(z^{t+1}\right)^2+1}{y_1y_2^2w^t}.
\end{align*}
It is easy to check that this is the QRT map $\varphi_{\rm CA}$ given by the matrices  (\ref{eq:CAQRTmat}) and the invariant curve is given by the formula (\ref{eq:invariantcurve}) as (\ref{eq:CAQRTinvari}).
Hence, by the successive application of the QRT map $\varphi_{\rm CA}$ to the point $(z^0,w^0)$ on the plane $\P^1(\C)\times\P^1(\C)$, we obtain a sequence $\left\{\left(z^t,w^t\right)\right\}_{t\geq0}$ of points on the invariant curve (\ref{eq:CAQRTinvari}) of the map $\varphi_{\rm CA}$.
\qed

The above procedure reducing the QRT map $\varphi_{\rm CA}$ from the seed mutations $\mu_1$ and $\mu_2$ in the cluster algebra $\mathcal{A}$ of type $A^{(1)}_1$ can be applied to any cluster algebra of rank 2.
We will list all map dynamical systems reduced from seed mutations in the rank 2 cluster algebras of finite and affine types in appendix.

It is well known that the cluster variables in any cluster algebra are Laurent polynomials in the initial cluster variables.
We refer to this property as the Laurent phenomenon \cite{FZ03}.
By using the invariant curve (\ref{eq:CAQRTinvari}) of $\varphi_{\rm CA}$, we obtain an invariant of the seed mutations in the cluster algebra $\mathcal{A}$ which exhibits the Laurent phenomenon.

\begin{proposition}\label{prop:CAmutinv}
Let $\mathcal{A}$ be the cluster algebra of type $A^{(1)}_1$.
Also let $\T_2\ni t_m\mapsto\Sigma_m$ be the cluster pattern of $\mathcal{A}$.
Then the following
\begin{align*}
\frac{y_1y_2^2(x_{2;n})^2+y_1^2y_2^3(x_{1;n})^2+(y_1y_2)^{2n}}{x_{1;n}x_{2;n}}
\end{align*}
is the Laurent polynomial in the initial cluster variables $x_1$ and $x_2$ at any vertex $t_n\in\T_2$:
\begin{align*}
\frac{y_1y_2^2(x_{2})^2+y_1^2y_2^3(x_{1})^2+1}{x_{1}x_{2}}.
\end{align*}
\end{proposition}

(Proof)\qquad
Since (\ref{eq:CAQRTinvari}) is the invariant curve of the QRT map $\varphi_{\rm CA}$, the conserved quantity $\lambda$ of $\varphi_{\rm CA}$ is given by the Laurent polynomial in $z$ and $w$:
\begin{align*}
\lambda=
-\frac{y_1y_2^2w^2+y_1^2y_2^3z^2+1}{zw}.
\end{align*}
The statement is obvious by virtue of this fact.
\qed

Geometrically, a QRT map is an addition of points on its invariant curve when it is an elliptic curve.
Unfortunately, the invariant curve (\ref{eq:CAQRTinvari}) of the QRT map $\varphi_{\rm CA}$ is not an elliptic curve but a conic.
In the following section, we show that the QRT map $\varphi_{\rm CA}$ is a degenerate limit of the QRT map $\varphi_{\rm TL}$ arising from the discrete Toda lattice $\chi_1$ of type $A^{(1)}_1$ and how the elliptic curve $\widetilde\gamma_1$ arising as the spectral curve of $\chi_1$ degenerates into the quadratic curve (\ref{eq:CAQRTinvari}).

\section{From discrete Toda  lattice to cluster algebra via QRT maps}
\label{sec:TLCA}
Let us observe the two sets of matrices (\ref{eq:TodaQRTmat}) and (\ref{eq:CAQRTmat}), the former is arising from the discrete Toda lattice and the latter from the cluster algebra both of which are of type $A^{(1)}_1$. 
We then find that if $b=0$ the matrices $B$'s coincide with each other, while $A$'s are a bit different:
\begin{align*}
A_{\rm TL}
=
\left(\begin{matrix}
0&0&a\\
0&0&1\\
1&0&0\\
\end{matrix}\right),\qquad
A_{\rm CA}
=
\left(\begin{matrix}
0&0&y_1^2y_2^3\\
0&0&0\\
y_1y_2^2&0&1\\
\end{matrix}\right).
\end{align*}

For the QRT maps $\varphi_{\rm TL}$ and $\varphi_{\rm CA}$ respectively associated with $\left\{A_{\rm TL},B_{\rm TL}\right\}$ and $\left\{A_{\rm CA},B_{\rm CA}\right\}$, an appropriate choice of $y_1$ and $y_2$ relates them.
\begin{theorem}\label{thm:TLCA}
Suppose $\xi,\eta,a\in\C$ and
\begin{align}
b=0.
\label{eq:condb}
\end{align}
Also suppose
\begin{align}
y_1=a^2\xi
\qquad\mbox{and}\qquad
y_2=1/a\xi.
\label{eq:condy}
\end{align}
Then we have
\begin{align}
\xi\varphi_{\rm TL}^{2n}(\xi,\eta)=x^{(n)}\varphi_{\rm CA}^{n}(\xi,\eta)
\label{eq:phimurel}
\end{align}
for $n\geq0$, where $x^{(n)}$ is the $x$-component of $\varphi_{\rm CA}^{n}(\xi,\eta)$.
\end{theorem}

(Proof)\qquad
Suppose $b=0$.
Then the QRT map $\varphi_{\rm TL}:(x^t,y^t)\mapsto(x^{t+1},y^{t+1})$ arising from the discrete Toda lattice of type $A^{(1)}_1$ has the following form
\begin{align*}
x^{t+1}
=
\frac{\left(y^{t}\right)^2}{ax^{t}},\qquad
y^{t+1}
=
\frac{a\left(x^{t+1}\right)^2+x^{t+1}}{y^{t}}.
\end{align*}
The iteration $\varphi_{\rm TL}^2:(x^t,y^t)\mapsto(x^{t+2},y^{t+2})$  of $\varphi_{\rm TL}$ leads to
\begin{align}
x^{t+2}
=
\frac{1}{x^t}\left(\frac{\left(y^{t}\right)^2+x^t}{ax^{t}}\right)^2,\qquad
y^{t+2}
=
\frac{a\left(x^{t}\right)^2x^{t+2}\left(ax^{t+2}+1\right)}{y^{t}\left(\left(y^t\right)^2+x^t\right)}.
\label{eq:TodaQRTdouble}
\end{align}

Let $x^{(n)}$ and $y^{(n)}$ be the $x$ and $y$-components of $\varphi_{\rm CA}^{n}(\xi,\eta)$, respectively.
Also let $x^{\{n\}}$ and $y^{\{n\}}$ be the $x$ and $y$-components of $\varphi_{\rm TL}^{n}(\xi,\eta)$, respectively.

We show (\ref{eq:phimurel}) in terms of induction on $n$.
Substituting $t=0$, $x^{\{0\}}=\xi$ and $y^{\{0\}}=\eta$ into (\ref{eq:TodaQRTdouble}), we have
\begin{align*}
\xi x^{\{2\}}
=
\left(\frac{\eta^2+\xi}{a\xi}\right)^2
\end{align*}
and
\begin{align*}
\xi y^{\{2\}}
=
\frac{a\xi^3x^{\{2\}}\left(ax^{\{2\}}+1\right)}{\eta\left(\eta^2+\xi\right)}
=
\frac{\eta^2+\xi}{a\xi}\times
\frac{\left(\eta^2+\xi\right)^2+a\xi^3}{a\xi^2\eta}.
\end{align*}
Similarly, for the QRT map $\varphi_{\rm CA}: (x^{(0)},y^{(0)})\mapsto(x^{(1)},y^{(1)})$ arising from the seed mutations in the cluster algebra $\mathcal{A}$ of type $A^{(1)}_1$, we have
\begin{align*}
x^{(1)}
&=
\frac{\left(y^{(0)}\right)^2+y_1^{-1}y_2^{-2}}{y_1y_2x^{(0)}}
=
\frac{\eta^2+\xi}{a\xi},\\
y^{(1)}
&=
\frac{y_1y_2\left(x^{(1)}\right)^2+y_1^{-1}y_2^{-2}}{y^{(0)}}
=
\frac{a\left(x^{(1)}\right)^2+\xi}{\eta}
=
\frac{\left(\eta^2+\xi\right)^2+a\xi^3}{a\xi^2\eta}.
\end{align*}
It immediately follows
\begin{align*}
\xi\varphi_{\rm TL}^{2}(\xi,\eta)=x^{(1)}\varphi_{\rm CA}(\xi,\eta).
\end{align*}

Assume that (\ref{eq:phimurel}) is true for $n$:
\begin{align*}
\xi x^{\{2n\}}=\left(x^{(n)}\right)^2,\qquad
\xi y^{\{2n\}}=x^{(n)}y^{(n)}.
\end{align*}
We then compute
\begin{align*}
\xi x^{\{2n+2\}}
&=
\frac{\xi}{x^{\{2n\}}}\left(\frac{\left(y^{\{2n\}}\right)^2+x^{\{2n\}}}{ax^{\{2n\}}}\right)^2\\
&=
\frac{\xi}{\frac{\left(x^{(n)}\right)^2}{\xi}}\frac{\left(\frac{x^{(n)}y^{(n)}}{\xi}\right)^4+2\left(\frac{x^{(n)}y^{(n)}}{\xi}\right)^2\frac{\left(x^{(n)}\right)^2}{\xi}+\left(\frac{\left(x^{(n)}\right)^2}{\xi}\right)^2}{a^2\left(\frac{\left(x^{(n)}\right)^2}{\xi}\right)^2}\\
&=
\left(
\frac{\left(y^{(n)}\right)^2+\xi}{a\left(x^{(n)}\right)}
\right)^2
=
\left(x^{(n+1)}\right)^2.
\end{align*}
Also we have
\begin{align*}
\xi y^{\{2n+2\}}
&=
\frac{\xi a\left(x^{\{2n\}}\right)^2x^{\{2n+2\}}\left(ax^{\{2n+2\}}+1\right)}{y^{\{2n\}}\left(\left(y^{\{2n\}}\right)^2+x^{\{2n\}}\right)}\\
&=
\frac{\xi a\frac{\left(x^{(n)}\right)^4}{\xi^2}\frac{\left(x^{(n+1)}\right)^2}{\xi}\left(a\frac{\left(x^{(n+1)}\right)^2}{\xi}+1\right)}{\frac{x^{(n)}y^{(n)}}{\xi}\left\{\left(\frac{x^{(n)}y^{(n)}}{\xi}\right)^2+\frac{\left(x^{(n)}\right)^2}{\xi}\right\}}\\
&=
x^{(n+1)}\times\frac{ax^{(n)}x^{(n+1)}}{\left(y^{(n)}\right)^2+\xi}\times
\frac{a\left(x^{(n+1)}\right)^2+\xi}{y^{(n)}}
=
x^{(n+1)}y^{(n+1)}.
\end{align*}
Induction on $n$ completes the proof.
\qed

By virtue of theorem \ref{thm:TLCA}, the clusters $\bx_m$ in the cluster algebra $\mathcal{A}$ of type $A^{(1)}_1$ are explicitly given by using the dependent variables $\bI^t$ and $\bV^t$ of the discrete Toda lattice $\chi_1$ of type $A^{(1)}_1$.
\begin{corollary}\label{cor:TLCAdirect}
Let $\mathcal{A}$ be the cluster algebra of type $A^{(1)}_1$ with imposing the initial cluster variable $x_1$ to be 1.
Also let $\T_2\ni t_m\mapsto\Sigma_m$ be the cluster pattern of $\mathcal{A}$.
Suppose that the solutions $I_2^{2n}$ and $V_2^{2n}$ to the discrete Toda lattice $\chi_1$ of type $A^{(1)}_1$ have the initial values
\begin{align}
I^0_1=\frac{x_1}{x_2}=\frac{1}{x_2},\qquad 
I^0_2=\frac{y_1y_2x_2}{x_1}=y_1y_2x_2,\qquad 
V^0_1=0,\qquad 
V^0_2=\frac{y_2}{x_2}.
\label{eq:TLiv}
\end{align}
Then the cluster $\bx_m=\left(x_{1;m},x_{2;m}\right)$ for $m\in\Z_{>0}$ is given as follows
\begin{align*}
x_{1;m}
=
\sqrt{\frac{\left(I_1^0I_2^0\right)^{2\overline{m}-1}}{I_2^{2\overline{m}}V_2^{2\overline{m}}}},
\qquad
x_{2;m}
=
\sqrt{\frac{\left(I_1^0I_2^0\right)^{2\underline{m}-1}I_2^{2\underline{m}}}{V_2^{2\underline{m}}}},
\end{align*} 
where $\overline{m}:=\left\lceil \frac{m}{2}\right\rceil=\min\left\{n\in\Z\ |\ n\geq\frac{m}{2}\right\}$ and $\underline{m}:=\left\lfloor \frac{m}{2}\right\rfloor=\max\left\{n\in\Z\ |\ n\leq\frac{m}{2}\right\}$.
Here the square roots are so chosen as to satisfy
\begin{align*}
x_{1;2n}x_{2;2n}
=
\frac{\left(I_1^0I_2^0\right)^{2n-1}}{V_2^{2n}}.
\end{align*}
\end{corollary}

(Proof)\qquad
The initial values (\ref{eq:TLiv}) lead to $b=-V_1^0V_2^0=0$, which satisfies the condition (\ref{eq:condb}) on $b$ in theorem \ref{thm:TLCA}.
Moreover, since $V_1^0=0$,  we see from (\ref{eq:TETL2}) that $V_1^n=0$ holds for $n\geq0$.

Noting (\ref{eq:xyIV}), the initial value $\left(x^{\{0\}},y^{\{0\}}\right)$ of the QRT map $\varphi_{\rm TL}$ arising from the Toda lattice $\chi_1$ is given as follows
\begin{align*}
x^{\{0\}}=\frac{1}{V_2^0\left(I_2^0+V_1^0\right)}=\frac{1}{I_2^0V_2^0}=x_1^{(0)},\qquad
y^{\{0\}}=\frac{1}{V_2^0}=x_2^{(0)},
\end{align*}
where $(x_1^{(0)},x_2^{(0)})=(x_1/y_1(y_2)^2,x_2/y_2)$ is the initial value of the QRT map $\varphi_{\rm CA}$ arising from the mutations of type $A^{(1)}_1$.
Thus, by theorem \ref{thm:TLCA}, we have
\begin{align*}
\left(x^{(n)}\right)^2
&=
\left(\frac{x_{1;2n}}{\left(y_1y_2\right)^{n}}\right)^2
=
x_1^{(0)}x^{\{2n\}}
=
\frac{x_1^{(0)}}{I_2^{2n}V_2^{2n}},\\
x^{(n)}y^{(n)}
&=
\frac{x_{1;2n}}{\left(y_1y_2\right)^{n}}\frac{x_{2;2n}}{\left(y_1y_2\right)^{n}}
=
x_1^{(0)}y^{\{2n\}}
=
\frac{x_1^{(0)}}{V_2^{2n}},
\end{align*}
where we use $V_1^{2n}=0$ and (\ref{eq:xyIV}) again.
It immediately follows
\begin{align*}
\left(\frac{x_{2;2n}}{\left(y_1y_2\right)^{n}}\right)^2
=
\frac{x_1^{(0)}I_2^{2n}}{V_2^{2n}}.
\end{align*}
The relations
\begin{align*}
x_1^{(0)}=\frac{1}{I_2^0V_2^0},\qquad
y_1y_2=I_1^0I_2^0,\qquad
x_{1;2n}=x_{1;2n-1},\qquad
x_{2;2n}=x_{2;2n+1}
\end{align*}
leads to the conclusion.
\qed

The following corollary to theorem \ref{thm:TLCA} is similarly shown.

\begin{corollary}\label{cor:TLCAdirect2}
Let $\T_2\ni t_m\mapsto\Sigma_m=(\bx_m,\by_m,B_m)$ be the cluster pattern of the cluster algebra $\mathcal{A}$ of type $A^{(1)}_1$ for the initial seed $\Sigma_0=\left(\bx_0,\by_0,B_0\right)$
\begin{align*}
\bx_0=(x_1,x_2)=\left(1,\frac{1}{I_1^0}\right),\qquad
\by_0=\left(y_1,y_2\right)=\left(y_1,\frac{I_1^0I_2^0}{y_1}\right),\qquad
B_0=\left(\begin{matrix}0&-2\\2&0\\\end{matrix}\right).
\end{align*}
Assume $V_1^0=0$.
Then the solution $\left(\bI^{2n},\bV^{2n}\right)$ to $\chi_1$  is given as
\begin{align*}
I_1^{2n}=y_1y_2\frac{x_{1;2n}}{x_{2;2n}},\qquad
I_2^{2n}=\frac{x_{2;2n}}{x_{1;2n}},\qquad
V_1^{2n}=0,\qquad
V_2^{2n}=\frac{(y_1)^{2n-1}(y_2)^{2n-2}}{x_{1;2n}x_{2;2n}}
\end{align*}
by using the cluster $\bx_{2n}$ in $\mathcal{A}$.
\qed
\end{corollary}

If we impose the condition (\ref{eq:condb}) on the invariant curve $\widetilde\gamma$ of the QRT map $\varphi_{\rm TL}$ then it reduces to the conic
\begin{align}
y^2+\lambda xy+ax^2+x=0.
\label{eq:TLinvarired}
\end{align}
The $2n$-th solution $(x^{\{2n\}},y^{\{2n\}})$ to $\varphi_{\rm TL}$, which is equivalent to $\left((x^{(n)})^2/x^{\{0\}},x^{(n)}y^{(n)}/x^{\{0\}}\right)$ by theorem \ref{thm:TLCA}, solves (\ref{eq:TLinvarired}), where $(x^{(n)},y^{(n)})$ is the $n$-th solution to the QRT map $\varphi_{\rm CA}$.
By substituting
\begin{align*}
(x,y)=\left((x^{(n)})^2/x^{\{0\}},x^{(n)}y^{(n)}/x^{\{0\}}\right)
\end{align*}
into (\ref{eq:TLinvarired}), we have
\begin{align*}
\left(y^{(n)}\right)^2+\lambda x^{(n)}y^{(n)}+a\left(x^{(n)}\right)^2+x^{\{0\}}=0
\end{align*}
for $x^{\{0\}}\neq0$.
Since the invariant curve (\ref{eq:CAQRTinvari}) of the QRT map $\varphi_{\rm CA}$ imposing the condition (\ref{eq:condy}) reduces to the conic
\begin{align*}
w^2+\lambda\xi zw+az^2+\xi=0,
\end{align*}
$(z,w)=(x^{(n)},y^{(n)})$ solves it.
Thus, we see that the invariant curve (\ref{eq:CAQRTinvari}) of $\varphi_{\rm CA}$ is a degeneration of the invariant curve $\widetilde\gamma$ of $\varphi_{\rm TL}$.
Moreover, addition  (\ref{eq:addontgam}) of points on $\widetilde\gamma$ induces the map $\varphi_{\rm CA}$ in the limit $b\to0$.

\section{Concluding remarks}
\label{sec:CONCL}
We gave a new expression of the periodic discrete Toda lattice $\chi_1$ of dimension 4 as the QRT map $\varphi_{\rm TL}$ by using the additive group structure of the elliptic curve $\gamma_1$ arising as its spectral curve.
In this formulation, we transform the spectral curve $\gamma_1$ of the Toda lattice $\chi_1$ into the bi-quadratic curve $\widetilde\gamma_1$ so that its additive group structure coincides with that of the invariant curve of the QRT map $\varphi_{\rm TL}$.
We also realized the seed mutations in the cluster algebra $\mathcal{A}$ of type $A^{(1)}_1$ as a birational map on $\P^1(\C)\times\P^1(\C)$ and represented it as the QRT map $\varphi_{\rm CA}$.
We then related the two QRT maps  $\varphi_{\rm TL}$ and  $\varphi_{\rm CA}$ directly by specializing the parameters contained in them and by choosing appropriate initial values.
This formulation gives a geometric interpretation of the seed mutations in $\mathcal{A}$ as a degeneration of the addition of points on the elliptic curve $\widetilde\gamma_1$.
Using the direct connection, cluster variables in $\mathcal{A}$ was given in terms of the solutions to the Toda lattice $\chi_1$ starting from appropriate initial variables, and vice versa.

Although we showed certain links of cluster algebras of rank 2 and QRT maps only in this paper, such phenomena can be found in cluster algebras of any rank because a seed mutation is essentially local, {\it i.e.}, it depends on an exchange relation among neighboring cluster variables linked by the exchange matrix.
Therefore, if we focus on two cluster variables, say $x_k$ and $x_l$, appearing in an exchange relation and consider the composition of the seed mutations in the directions $k$ and $l$ then it can be regarded as a map dynamical system on the plane, which is the intersection of the hyperplanes given by the unmutated variables.
By construction, a map dynamical system thus obtained depends only on the topology of the quiver associated with the exchange matrix.
Moreover, successive application of seed mutations induces a global map dynamical system which is the composition of the local ones.
We will report map dynamical systems arising from seed mutations in cluster algebras of general rank in a forthcoming paper.

\section*{Acknowledgments}
This work was partially supported by JSPS KAKENHI Grant Number 26400107.

\appendix
\section{Cluster algebras of rank 2 and QRT maps}
\label{asec:CAR2QRT}
Let us consider the cluster algebra of type $A_2$ whose initial seed $\Sigma_0=(\bx_0,\by_0,B_0)$ is given as follows
\begin{align*}
\bx_0=(x_1,x_2),\qquad
\by_0=(y_1,y_2),\qquad
B_0=\left(\begin{matrix}
0&-1\\
1&0\\
\end{matrix}\right).
\end{align*}
Note that we do not assume the semifield $\P$ to be tropical.
The Cartan counterpart $A(B_0)$ of $B_0$ is of type $A_2$:
\begin{align*}
A(B_0)
=\left(\begin{matrix}
2&-1\\
-1&2\\
\end{matrix}\right).
\end{align*}
The Dynkin diagram and the quiver associated with $B_0$ is given as follows
\begin{align*}
{\unitlength=.06in{\def\arraystretch{1.0}
\begin{picture}(12,5)(0,-.5)
\thicklines
\put(1,.5){\circle{2}}
\put(10,.5){\circle{2}}
\put(2,.5){\line(1,0){7}}
\put(1,3){\makebox(0,0){$1$}}
\put(10,3){\makebox(0,0){$2$}}
\end{picture},\qquad\qquad
\begin{picture}(12,5)(0,-.5)
\thicklines
\put(1,.5){\circle{2}}
\put(10,.5){\circle{2}}
\put(9,.5){\vector(-1,0){7}}
\put(1,3){\makebox(0,0){$1$}}
\put(10,3){\makebox(0,0){$2$}}
\end{picture}
}}.
\end{align*}

Let $\T_2\ni t_n\mapsto\Sigma_n=(\bx_n,\by_n,B_n)$ be the cluster pattern of the cluster algebra.
Consider the map $\bx_n\overset{\mu_1}{\longmapsto}\bx_{n+1}\overset{\mu_2}{\longmapsto}\bx_{n+2}$.
We then have
\begin{align*}
x_{1;n+2}
&=
x_{1;n+1}
=
\frac{y_{1;n}x_{2;n}+1}{(y_{1;n}\oplus1)x_{1;n}}
=
y_{2;n+2}\frac{y_{1;n}x_{2;n}+1}{\frac{x_{1;n}}{y_{2;n}}},\\
x_{2;n+2}
&=
\frac{y_{2;n+1}x_{1;n+1}+1}{(y_{2;n+1}\oplus1)x_{2;n+1}}
=
\frac{y_{2;n+1}x_{1;n+2}+1}{(y_{2;n+1}\oplus1)x_{2;n}}
=
\frac{1}{y_{1;n+2}}\frac{\frac{x_{1;n+2}}{y_{2;n+2}}+1}{y_{1;n}x_{2;n}},
\end{align*}
where we use $x_{1;n+1}=x_{1;n+2}$ , $x_{2;n}=x_{2;n+1}$ and the following relations obtained from the exchange relation (\ref{eq:mutcoef})  of $\by$:
\begin{align*}
y_{1;n+2}&=y_{1;n+1}(y_{2;n+1}\oplus1),\\
y_{1;n}y_{1;n+1}&=1,\\
y_{2;n+1}&=y_{2;n}(y_{1;n}\oplus1),\\
y_{2;n+1}y_{2;n+2}&=1.
\end{align*}
If we put
\begin{align*}
z^t=\frac{x_{1;2t}}{y_{2;2t}},\qquad
w^t=y_{1;2t}x_{2;2t}
\end{align*}
then we have
\begin{align}
z^{t+1}
=
\frac{w^t+1}{z^t},\qquad
w^{t+1}
=
\frac{z^{t+1}+1}{w^t}.
\label{eq:A2QRT}
\end{align}
We see that the map $(z^t,w^t)\mapsto(z^{t+1},w^{t+1})$ is the QRT map given by the matrices
\begin{align*}
A=\left(
\begin{matrix}
0&1&1\\
1&0&2\\
1&2&1\\
\end{matrix}
\right),\qquad
B=\left(
\begin{matrix}
0&0&0\\
0&1&0\\
0&0&0\\
\end{matrix}
\right).
\end{align*}
The invariant curve of the QRT map is the following cubic curve
\begin{align*}
(z+1)w^2+(z^2+\lambda z+2)w+(z+1)^2=0,
\end{align*}
and $\lambda$ is the conserved quantity of the map.

The QRT map (\ref{eq:A2QRT}) has period 5 for a generic initial point, which reflects the fact that the corresponding cluster algebra is of finite type having 5 independent cluster variables.

In the same manner, we obtain birational-map dynamical systems on $\P^1(\C)\times\P^1(\C)$ from the cluster algebras of types $A_1\times A_1$, $B_2$, $G_2$, $A^{(1)}_1$ and $A^{(2)}_2$, which complete the cluster algebras of rank 2 associated with finite and affine Lie algebras.
Three of them ($A_1\times A_1$, $B_2$ and $A^{(1)}_1$) lead to QRT maps, while $G_2$ and $A^{(2)}_2$ induce dynamical systems of higher degree.
We list them in the following.
We use the same notations as in the case of type $A_2$ unless otherwise stated.

\begin{itemize}
\item Type $A_1\times A_1$

Initial exchange matrix and its Cartan counterpart:
\begin{align*}
B_0=\left(\begin{matrix}
0&0\\
0&0\\
\end{matrix}\right),\qquad
A(B_0)
=\left(\begin{matrix}
2&0\\
0&2\\
\end{matrix}\right).
\end{align*}
Dynkin diagram and quiver associated with $B_0$:
\begin{align*}
{\unitlength=.06in{\def\arraystretch{1.0}
\begin{picture}(12,5)(0,0)
\thicklines
\put(1,1){\circle{2}}
\put(10,1){\circle{2}}
\put(1,3.5){\makebox(0,0){$1$}}
\put(10,3.5){\makebox(0,0){$2$}}
\end{picture},
\qquad\qquad
\begin{picture}(12,5)(0,0)
\thicklines
\put(1,1){\circle{2}}
\put(10,1){\circle{2}}
\put(1,3.5){\makebox(0,0){$1$}}
\put(10,3.5){\makebox(0,0){$2$}}
\end{picture}
}}.
\end{align*}
Variable transformation:
\begin{align*}
z^t=\sqrt{\frac{1+y_1}{1\oplus y_1}}x_{1;2t},\qquad
w^t=\sqrt{\frac{1+y_2}{1\oplus y_2}}x_{2;2t}.
\end{align*}
QRT map and its matrices:
\begin{align*}
&z^{t+1}
=
\frac{1}{z^t},\qquad
w^{t+1}
=
\frac{1}{w^t},\\
&A=\left(
\begin{matrix}
0&1&0\\
1&0&1\\
0&1&0\\
\end{matrix}
\right),\qquad
B=\left(
\begin{matrix}
0&0&0\\
0&1&0\\
0&0&0\\
\end{matrix}
\right).
\end{align*}
Invariant curve:
\begin{align*}
zw^2+z^2w+\lambda zw+w+z=0.
\end{align*}

\item Type $B_2$

Initial exchange matrix and its Cartan counterpart:
\begin{align*}
B_0=\left(\begin{matrix}
0&-2\\
1&0\\
\end{matrix}\right),\qquad
A(B_0)
=\left(\begin{matrix}
2&-2\\
-1&2\\
\end{matrix}\right).
\end{align*}
Dynkin diagram (since $B_0$ is not skew-symmetric, there is no quiver associated with $B_0$):
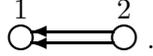
\begin{align*}
{\unitlength=.06in{\def\arraystretch{1.0}
\begin{picture}(12,5)(0,0)
\thicklines
\put(1,1){\circle{2}}
\put(10,1){\circle{2}}
\put(9.1,1.5){\vector(-1,0){7.2}}
\put(9.1,.5){\vector(-1,0){7.2}}
\put(1,3.5){\makebox(0,0){$1$}}
\put(10,3.5){\makebox(0,0){$2$}}
\end{picture}
}}.
\end{align*}
Variable transformation:
\begin{align*}
z^t=\frac{x_{1;2t}}{\sqrt{y_{2;2t}}},\qquad
w^t=y_{1;2t}x_{2;2t}.
\end{align*}
QRT map and its matrices:
\begin{align*}
&z^{t+1}
=
\frac{w^t+1}{z^t},\qquad
w^{t+1}
=
\frac{\left(z^{t+1}\right)^2+1}{w^t},\\
&A=\left(
\begin{matrix}
0&1&1\\
0&0&0\\
1&2&1\\
\end{matrix}
\right),\qquad
B=\left(
\begin{matrix}
0&0&0\\
0&1&0\\
0&0&0\\
\end{matrix}
\right).
\end{align*}
Invariant curve:
\begin{align*}
w^2+(z^2+\lambda z+2)w+z^2+1=0.
\end{align*}

\item Type $G_2$

Initial exchange matrix and its Cartan counterpart:
\begin{align*}
B_0=\left(\begin{matrix}
0&1\\
-3&0\\
\end{matrix}\right),\qquad
A(B_0)
=\left(\begin{matrix}
2&-1\\
-3&2\\
\end{matrix}\right).
\end{align*}
Dynkin diagram:
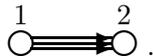
\begin{align*}
{\unitlength=.06in{\def\arraystretch{1.0}
\begin{picture}(12,5)(0,0)
\thicklines
\put(1,1){\circle{2}}
\put(10,1){\circle{2}}
\put(1.9,1.5){\vector(1,0){7.1}}
\put(2,1){\vector(1,0){7}}
\put(1.9,.5){\vector(1,0){7.1}}
\put(1,3.5){\makebox(0,0){$1$}}
\put(10,3.5){\makebox(0,0){$2$}}
\end{picture}
}}.
\end{align*}
Variable transformation:
\begin{align*}
z^t=y_{2;2t}x_{1;2t},\qquad
w^t=\frac{x_{2;2t}}{\sqrt[3]{y_{1;2t}}}.
\end{align*}
Map dynamical system (not a QRT map):
\begin{align*}
z^{t+1}
=
\frac{\left(w^t\right)^3+1}{z^t},\qquad
w^{t+1}
=
\frac{z^{t+1}+1}{w^t}.
\end{align*}
Invariant curve:
\begin{align*}
z&+w+\frac{1+z}{w}+\frac{(1+z)^3+w^3}{zw^3}+\frac{(1+z)^2+w^3}{zw^2}\\
&+\frac{(1+z)^3+2w^3+w^6+3zw^3}{z^2w^3}+\frac{1+z+w^3}{zw}+\frac{1+w^3}{z}=\lambda.
\end{align*}
\begin{remark}
Since the cluster algebra of type $G_2$ is of finite type having period 8, the sum of all 8 cluster variables is invariant under the seed mutation.
This gives the above sextic invariant curve \cite{HY02}.
\end{remark}

\item Type $A^{(1)}_1$

Initial exchange matrix and its Cartan counterpart:
\begin{align*}
B_0=\left(\begin{matrix}
0&-2\\
2&0\\
\end{matrix}\right),\qquad
A(B_0)
=\left(\begin{matrix}
2&-2\\
-2&2\\
\end{matrix}\right).
\end{align*}
Dynkin diagram and quiver associated with $B_0$:
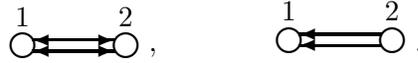
\begin{align*}
{\unitlength=.06in{\def\arraystretch{1.0}
\begin{picture}(12,5)(0,0)
\thicklines
\put(1,.5){\circle{2}}
\put(10,.5){\circle{2}}
\put(1.8,1){\vector(1,0){7.2}}
\put(1.8,0){\vector(1,0){7.2}}
\put(2.4,1){\vector(-1,0){.5}}
\put(2.4,0){\vector(-1,0){.5}}
\put(1,3){\makebox(0,0){$1$}}
\put(10,3){\makebox(0,0){$2$}}
\end{picture},
\qquad\qquad
\begin{picture}(12,5)(0,0)
\thicklines
\put(1,1){\circle{2}}
\put(10,1){\circle{2}}
\put(9.1,1.5){\vector(-1,0){7.2}}
\put(9.1,.5){\vector(-1,0){7.2}}
\put(1,3.5){\makebox(0,0){$1$}}
\put(10,3.5){\makebox(0,0){$2$}}
\end{picture}
}}.
\end{align*}
Variable transformation:
\begin{align*}
z^t=\frac{x_{1;2t}}{\sqrt{y_{2;2t}}},\qquad
w^t=\sqrt{y_{1;2t}}x_{2;2t}.
\end{align*}
QRT map and its matrices:
\begin{align}
&z^{t+1}
=
\frac{(w^t)^2+1}{z^t},\qquad
w^{t+1}
=
\frac{\left(z^{t+1}\right)^2+1}{w^t},\nonumber\\
&A=\left(
\begin{matrix}
0&0&1\\
0&0&0\\
1&0&1\\
\end{matrix}
\right),\qquad
B=\left(
\begin{matrix}
0&0&0\\
0&1&0\\
0&0&0\\
\end{matrix}
\right).\label{eq:A11general}
\end{align}
Invariant curve:
\begin{align*}
w^2+\lambda zw+z^2+1=0.
\end{align*}

\begin{remark}
The QRT map (\ref{eq:A11general}) is independent of the choice of $\P$.
Therefore, if we set $y_1=y_2=1$ in the tropical case (\ref{eq:CAQRTmat}) we obtain  (\ref{eq:A11general}).
\end{remark}

\item Type $A^{(2)}_2$

Initial exchange matrix and its Cartan counterpart:
\begin{align*}
B_0=\left(\begin{matrix}
0&-4\\
1&0\\
\end{matrix}\right),\qquad
A(B_0)
=\left(\begin{matrix}
2&-4\\
-1&2\\
\end{matrix}\right).
\end{align*}
Dynkin diagram:
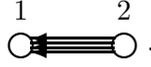
\begin{align*}
{\unitlength=.06in{\def\arraystretch{1.0}
\begin{picture}(12,5)(0,0)
\thicklines
\put(1,.5){\circle{2}}
\put(10,.5){\circle{2}}
\put(9.2,1.1){\vector(-1,0){7.3}}
\put(9.1,.7){\vector(-1,0){7.2}}
\put(9.1,.3){\vector(-1,0){7.2}}
\put(9.2,-.1){\vector(-1,0){7.3}}
\put(1,3.5){\makebox(0,0){$1$}}
\put(10,3.5){\makebox(0,0){$2$}}
\end{picture}.
}}
\end{align*}
Variable transformation:
\begin{align*}
z^t=\frac{x_{1;2t}}{\sqrt[4]{y_{2;2t}}},\qquad
w^t=y_{1;2t}x_{2;2t}.
\end{align*}
Map dynamical system (not a QRT map):
\begin{align*}
z^{t+1}
=
\frac{w^t+1}{z^t},\qquad
w^{t+1}
=
\frac{\left(z^{t+1}\right)^4+1}{w^t}.
\end{align*}
The invariant curve has not been obtained yet (since the cluster algebra of type $A^{(2)}_2$ is of infinite type, we can not apply the same procedure as in the case of type $G_2$).

\end{itemize}



\begin{thebibliography}{99}
\bibitem{FZ02}Fomin S and Zelevinsky A 2002 \textit{J. Amer. Math. Soc.} \textbf{15} 497-529
\bibitem{FZ03}Fomin S and Zelevinsky A 2003 \textit{Invent. Math.} \textbf{154} 63-121
\bibitem{FZ03-2}Fomin S and Zelevinsky A 2003 \textit{Ann. of Math.} \textbf{158} 977-1018
\bibitem{FZ06}Fomin S and Zelevinsky A 2007 \textit{Compos. Math.} \textbf{143} 112-64
\bibitem{Hirota77}Hirota R 1977 \textit{J. Phys. Soc. Jpn.} \textbf{43} 2074-8
\bibitem{HTI93} Hirota R, Tsujimoto S and Imai T 1993 ``Difference scheme of soliton equations" in \textit{Future Directions of Nonlinear Dynamics in Physical and Biological Systems} edited by Christiansen P L, Eilbeck J G and Parmentier R D (New York: Plenum)
\bibitem{HY02}Hirota R and Yahagi H 2002 \textit{J. Phys. Soc. Jpn.} \textbf{71} 2867-72
\bibitem{IIKNS10}Inoue R, Iyama O, Kuniba A, Nakanishi T and Suzuki J 2010 \textit{Nagoya Math. J.} \textbf{197} 59-174
\bibitem{IIKKN13}Inoue R, Iyama O, Keller B, Kuniba A and Nakanishi T 2013 \textit{Publ. RIMS} \textbf{49} 1-42
\bibitem{IIKKN13-2}Inoue R, Iyama O, Keller B, Kuniba A and Nakanishi T 2013 \textit{Publ. RIMS} \textbf{49} 43-85
\bibitem{Iwao08}Iwao S 2008 \textit{J. Phys. A: Math. Theor.} \textbf{41} 115201
\bibitem{Mase13}Mase T 2013 \textit{RIMS K\^oky\^uroku Bessatsu} \textbf{B41} 43-64
\bibitem{Mase16}Mase T 2016 \textit{J. Math. Phys.} \textbf{57} 022703
\bibitem{Nobe13} Nobe A 2013 \textit{J. Phys. A: Math. Theor.} \textbf{46} 465203
\bibitem{Okubo13}Okubo N 2013 \textit{RIMS K\^oky\^uroku Bessatsu} \textbf{B41} 25-42
\bibitem{QRT89}Quispel G R W, Roberts J A G and Thompson C J 1989 \textit{Physica D} \textbf{34} 183-92
\bibitem{Suris03}Suris Y B 2003 ``The Problem of Integrable Discretization: Hamiltonian Approach" (Basel: Birkh\"auser)
\bibitem{Tsuda04}Tsuda T 2004 \textit{J. Phys. A: Math. Gen.} \textbf{37} 2721-30
\end{thebibliography}
\end{document}